\documentclass{article}
\usepackage{amssymb}

\begin{document}

\title{ON\ THE\ RELATION\ BETWEEN\ THE\ FIRST\ AND\ SECOND\ MOMENTS\ OF\
DISTRIBUTIONS}
\author{\textit{H. M. YEHIA} \\
\textit{Department of Mathematics, Faculty of Science,}\\
\textit{\ Mansoura University, Mansoura 35516, Egypt}\\
E-mail: \textit{hyehia@mans.edu.eg}}
\maketitle

\begin{abstract}
A condition on the location of the centre of a mass (or probability)
distribution is found if its second moments are given. The result is applied
to the relation between the centre of mass and the inertia matrix of bodies.
An examples is given to illustrate the importance of this condition.
\end{abstract}

\begin{center}
\bigskip ------------ \ \ \ \ \ \ \ \ \ \ \ \ \ \ \ ------------ \ \ \ \ \ \
\ \ \ \ \ \ \ \ ------------
\end{center}

\bigskip This note was published as:

\bigskip Journal of Physics A: Mathematical and General. Volume 35 (2002),
No 30, pp. 6505-8, doi:10.1088/0305-4470/35/30/321.

\begin{center}
------------ \ \ \ \ \ \ \ \ \ \ \ \ \ \ \ ------------ \ \ \ \ \ \ \ \ \ \
\ \ \ \ ------------
\end{center}

It is well known that for three given quantities to be the moments of
inertia of a real mass distribution with respect to three orthogonal axes
intersecting at some point, they must satisfy the triangle inequalities.
When one of the triangle inequalities renders to an equality, the mass
distribution must be planar and its centre of mass will lie in that plane.
Those facts are mentioned in most textbooks on mechanics of rigid bodies
(see e.g. \cite{mcm}, \cite{rou}). However, the reader is usually left with
the impression that the centre of mass of the distribution with given
moments of inertia can be chosen in an arbitrary way, which is in fact
incorrect.

Let $\Phi (x_{1},...,x_{n})$ be a normed n-dimensional non-negative
(continuous or discrete) distribution function. Denote by $\overline{x_{i}}$ 
$(i=1,...,n)$ the first moments of the distribution with respect to the
origin. We have the following

\textbf{Theorem:} \textit{If the matrix of the second moments of the
distribution with respect to the origin }%
\begin{equation}
S=\left( s_{ij}=\overline{x_{i}x_{j}}\right) _{i,j=1}^{n}  \label{1}
\end{equation}%
\textit{is given, then its centre }$(\overline{x_{1}},...,\overline{x_{n}})$%
\textit{\ lies inside or on the ellipsoid }%
\begin{equation}
\left\vert 
\begin{array}{lllll}
1 & x_{1} & x_{2} & \cdots & x_{n} \\ 
x_{1} & s_{11} & s_{12} & \cdots & s_{1n} \\ 
x_{2} & s_{12} & s_{22} & \cdots & s_{2n} \\ 
\vdots & \vdots & \vdots &  & \vdots \\ 
x_{n} & s_{1n} & s_{2n} & \cdots & s_{nn}%
\end{array}%
\right\vert =0  \label{2}
\end{equation}%
\textit{whose principal axes coincide with the eigenvectors of }$S$\textit{\
and whose semiaxes are the square roots of the eigenvalues of }$S$\textit{.}

\textbf{Proof}: Consider the quadratic form 
\begin{eqnarray}
f(\alpha _{0},\alpha _{1},...,\alpha _{n}) &=&\overline{(\alpha _{0}+\alpha
_{1}x_{1}+...+\alpha _{n}x_{n})^{2}}  \nonumber \\
&=&\alpha _{0}^{2}+2\alpha _{0}\sum_{i=1}^{n}\alpha _{i}\overline{x_{i}}%
+\sum_{i,j=1}^{n}\alpha _{i}\alpha _{j}s_{ij}  \nonumber \\
&=&\mathbf{\alpha A\alpha \acute{}}  \label{3}
\end{eqnarray}%
where $\mathbf{\alpha }=(\alpha _{0},\alpha _{1},...,\alpha _{n})$, $\alpha 
\acute{}$ its transpose and $\mathbf{A}$ is the matrix 
\begin{equation}
\mathbf{A}=\left( 
\begin{array}{lllll}
1 & \overline{x_{1}} & \overline{x_{2}} & \cdots & \overline{x_{n}} \\ 
\overline{x_{1}} & s_{11} & s_{12} & \cdots & s_{1n} \\ 
\overline{x_{2}} & s_{12} & s_{22} & \cdots & s_{2n} \\ 
\vdots & \vdots & \vdots &  & \vdots \\ 
\overline{x_{n}} & s_{1n} & s_{2n} & \cdots & s_{nn}%
\end{array}%
\right)  \label{4}
\end{equation}%
The quadratic form (\ref{3}) is non-negative for all real $\{\alpha _{i}\},$
and hence $\mathbf{A}$ must have non-negative principal minors\cite{gant}: 
\begin{equation}
A\left( 
\begin{array}{lll}
i_{1} & \cdots & i_{m} \\ 
i_{1} & \cdots & i_{m}%
\end{array}%
\right) \geqslant 0,1\leq i_{1}<\cdots <i_{m}\leq n+1,1\leq m\leq n+1
\label{5}
\end{equation}

Turning the axes $x_1,...,x_n$ to coincide with the principal axes $\xi
_1,...,\xi _n$ of $\mathbf{S}$, we replace $\mathbf{A}$ by 
\begin{equation}
\left( 
\begin{array}{lllll}
1 & \overline{\xi _1} & \overline{\xi _2} & \cdots & \overline{\xi _n} \\ 
\overline{\xi _1} & s_{11} & 0 & \cdots & 0 \\ 
\overline{\xi _2} & 0 & s_{22} & \cdots & 0 \\ 
\vdots & \vdots & \vdots & \ddots & \vdots \\ 
\overline{\xi _n} & 0 & 0 & \cdots & s_{nn}%
\end{array}
\right)  \label{6}
\end{equation}
and conditions (\ref{5}) take the form 
\begin{equation}
\overline{\xi _i^2}\geqslant 0,\xi _{i_1}^2\cdots \xi _{i_m}^2(1-\frac{%
\overline{\xi _{i_1}}^2}{\overline{\xi _{i_1}^2}}-\cdots -\frac{\overline{%
\xi _{i_m}}^2}{\overline{\xi _{i_m}^2}})\geqslant 0  \label{7}
\end{equation}

Now we have two cases:

\begin{enumerate}
\item If $\mathbf{S}$ is non-singular, then the set of conditions (\ref{7})
has its intersection as 
\begin{equation}
\sum_{i=1}^n\frac{\overline{\xi _i}^2}{\overline{\xi _i^2}}\leq 1  \label{8}
\end{equation}
meaning that the centre of the distribution must lie inside or on the
ellipsoid with semi-axes $\{\sqrt{\overline{\xi _i^2}}\}$ which are directed
along the $\xi _1,...,\xi _n$ axes. Returning back to the axes $\{x_i\}$ the
equation of the ellipsoid takes the form stated in the theorem.

\qquad The condition (\ref{8}) is not known in the literature. A fact
well-known for distributions and widely used in textbooks of statistics and
quantum mechanics is that for each variable $\overline{\xi _{i}}^{2}\leq 
\overline{\xi _{i}^{2}},$ so that 
\begin{equation}
\frac{\overline{\xi _{i}}^{2}}{\overline{\xi _{i}^{2}}}\leq 1,i=1,...,n
\label{8a}
\end{equation}%
This condition is much weaker than (\ref{8}) and it signifies that the
centre of the distribution lies inside or on the cuboid with sides $\{2\sqrt{%
\overline{\xi _{i}^{2}}}\}$ formed by tangent planes to the ellipsoid in (%
\ref{8}) at the ends of its axes. As a measure for comparison let us find
the ratio of the volumes $v_{e},v_{c}$ of admissible regions for the centre
under conditions (\ref{8}) and (\ref{8a}), respectively 
\begin{equation}
\frac{v_{e}}{v_{c}}=\frac{\frac{\pi ^{n/2}}{\Gamma (1+\frac{n}{2})}%
\prod\limits_{i=1}^{n}\sqrt{\overline{\xi _{i}^{2}}}}{2^{n}\prod%
\limits_{i=1}^{n}\sqrt{\overline{\xi _{i}^{2}}}}=\frac{(\frac{\sqrt{\pi }}{2}%
)^{n}}{\Gamma (1+\frac{n}{2})}
\end{equation}%
This ratio is $<1$ for all $n>1$ and decays quickly with growing $n.$ Some
values are given in the following table

$%
\begin{tabular}{llllllll}
$n$ & $1$ & $2$ & $3$ & $4$ & $5$ & $10$ & $20$ \\ 
$v_e/v_c$ & $1$ & $0.7854$ & $0.5236$ & $.3084$ & $0.1645$ & $.0025$ & $2.5$ 
$10^{-8}$%
\end{tabular}
$

\item If $\mathbf{S}$ is singular, then some of its eigenvalues are zeroes,
say $\overline{\xi _{k+1}^2}=\cdots =\overline{\xi _n^2}=0.$ The part of the
conditions (\ref{7}) in which those quantities appear reduce to 
\[
\overline{-\xi _{k+1}^2}\geqslant 0,\cdots ,-\overline{\xi _n^2}\geqslant 0 
\]
and hence 
\begin{equation}
\overline{\xi _{k+1}}=\cdots =\overline{\xi _n}=0  \label{9}
\end{equation}
The rest of the conditions (\ref{7}) have the intersection 
\begin{equation}
\sum_{i=1}^k\frac{\overline{\xi _i}^2}{\overline{\xi _i^2}}\leq 1  \label{10}
\end{equation}
The centre of the distribution in this case lies inside or on an ellipsoid
in the $k$-dimensional subspace $\xi _{k+1}=\cdots =\xi _n=0.$
\end{enumerate}

\textbf{Application to the centre of mass of a body with given inertia
matrix:}

Let $Oxyz$ be the Cartesian coordinate system coinciding with the principal
axes of inertia of a certain body of mass $M$ and given principal moments of
inertia $A,B$ and $C$, respectively. According to the above theorem, the
centre of mass of the body lies inside or on the ellipsoid 
\begin{equation}
\frac{x^2}{a^2}+\frac{y^2}{b^2}+\frac{z^2}{c^2}=1
\end{equation}
where 
\begin{equation}
a^2=\frac{B+C-A}{2M},b^2=\frac{C+A-B}{2M},c^2=\frac{A+B-C}{2M}  \label{a}
\end{equation}

Although one can expect such result should be mentioned somewhere in
classical textbooks on Mechanics of rigid bodies such as \cite{rou} and \cite%
{mcm}, this is not true. This result is new.

Note that if $C\rightarrow A+B$ then $c\rightarrow 0$ and the centre of mass
lies in the ellipse 
\begin{equation}
\frac{x^2}{B/M}+\frac{y^2}{A/M}=1  \label{b}
\end{equation}
of the plane $z=0$.

Such considerations were utilized in our work \cite{yz} in the context of
stability analysis of certain periodic motions of a rigid body about a fixed
point. In some works in applied mechanics those conditions are usually
overlooked, leading to unrealistic choices of parameters of rigid bodies in
numerical examples. An example is the work \cite{ism}, where those
parameters were 
\[
A=10,B=20,C=30,M=300,x_0=2,y_0=5 
\]
It is obvious, according to (\ref{b}), that the values of $x_0,y_0$
compatible with the given mass and moments must satisfy the condition $\frac{%
x^2}{a^2}+\frac{y^2}{b^2}<1,$ where $a=0.2582$ and $b=0.1826.$ If one
insists to keep the given values of $x_0,y_0$ and modify only the mass $M,$
the suitable choice would satisfy $M<0.3704.$

\end{document}